\documentclass[structabstract]{aa}
\usepackage{natbib}
\usepackage{txfonts}
\usepackage{graphicx}

\begin{document}

\title{Is there a mass discrepancy in the Cepheid binary OGLE-LMC-CEP0227?}

\author{Hilding R. Neilson \and Norbert Langer}
\titlerunning{Cepheid binary and mass discrepancy}
\authorrunning{Neilson et al.}
\institute{Argelander-Institut f\"{u}r Astronomie, Universit\"{a}t Bonn, Auf dem H\"{u}gel 71, 53121 Bonn Germany}
\date{}

\abstract{The Cepheid mass discrepancy, the difference between masses predicted from stellar evolution and stellar pulsation calculations, is a challenge for the understanding of stellar astrophysics. Recent models of the eclipsing binary Cepheid OGLE-LMC-CEP-0227 have suggested that the discrepancy may be resolved.}
{We explore for what physical parameters do stellar evolution models agree with the measured properties of OGLE-LMC-CEP0227 and compare to canonical stellar evolution models  assuming no convective core overshooting.}
{We construct state-of-the-art stellar evolution models for varying mass, metallicity, and convective core overshooting and compare the stellar evolution predictions with the observed properties.}
{The observed mass, effective temperature, and radius of the two stars in the binary system are well fit by numerous combinations of physical parameters, suggesting a Cepheid mass discrepancy of 10-20\% relative to canonical stellar evolution models.}
{The properties of the observed binary Cepheid suggest that the Cepheid mass discrepancy is still a challenge and requires more specific observations, such as the rate of period change, to better constrain and understand the necessary physics for stellar evolution models to resolve the discrepancy.}


\maketitle

\section{Introduction}\label{sec:intro}
Classical Cepheids are ideal laboratories for stellar astrophysics and evolution.  Stellar pulsation provides a measure of the fundamental properties of these stars including luminosity, radius, and mass that can be compared to predictions based on stellar evolution models.  Cepheid pulsation masses, however, do often not agree with predictions from stellar evolution models \citep{Stobie1969, Cox1980, Keller2008}.

This Cepheid mass discrepancy, defined through the relative difference between evolutionary and pulsation masses, is currently measured to be about 17\% \citep{Keller2008} and may be a function of mass \citep{Caputo2005} and metallicity \citep{Keller2006}.  There have been a number of solutions suggested to solve this mass discrepancy including convective core overshooting in the main sequence progenitors of Cepheids and mass loss during the Cepheid stage of evolution \citep{Bono2006}. \cite{Neilson2011} showed that the mass discrepancy may be solved by the combination of pulsation-driven mass loss during the Cepheid stage of evolution and moderate convective core overshooting. All of these suggestions imply the origin of the mass discrepancy lies within the stellar evolution models.

Dynamically measured  masses from Cepheid binaries have the potential to constrain and verify stellar pulsation or stellar evolution masses. Dynamical masses have been determined for a handful of Galactic Cepheids \citep{Evans1990, Evans1997, Evans2006, Evans2008}, suggesting that dynamic masses are consistent with pulsation masses.  The recent analysis of the eclipsing binary OGLE-LMC-CEP0227 by \cite{Piet2010}, however, obtains a Cepheid mass of $M=4.14\pm 0.05~M_\odot$, the most precise Cepheid measurement to date. The authors also derive the stellar pulsation mass for this object to be $M_P = 3.98\pm 0.28~M_\odot$, consistent with the dynamic mass.  The authors argued that because the pulsation mass agreed with the dynamic mass then the Cepheid mass discrepancy can only be solved by new physics in the stellar evolution models. However, they did not test this assertion with comparisons to stellar evolution models.

\cite{Cassisi2011} considered stellar evolution models with moderate convective core overshooting and found that the models fit the observed effective temperature and radius for stellar masses consistent with dynamic masses.  As a result, the authors conclude that there is no Cepheid mass discrepancy, at least for this particular Cepheid.   It is debatable whether this is a reasonable conclusion.  On one hand there is significant evidence suggesting the need to include convective overshooting in evolution models \citep[e.g.][]{Sandberg2010, Lovekin2010}. 

On the other hand, the underlying physics of convective core overshooting is not understood, hence the physics must be parameterized. Furthermore, because overshooting is parameterized in evolution models, it is not necessarily treated in the same way by various evolution codes. Another issue is that various observations constraining overshooting seem to require differing amounts of overshooting \cite[e.g.][]{Keller2006, Lovekin2010, Sandberg2010}. Finally, for Cepheids in particular, parameterizations of overshooting may act to hide different physics.  For instance, \cite{Bono2006} noted that the Cepheid mass discrepancy might be resolved by rotational mixing, mass loss, unknown opacities as well as overshooting. These different physical processes are degenerate when predicting stellar masses. 

 \cite{Keller2008} and \cite{Neilson2011} predict a similar amount of Cepheid mass discrepancy by assuming different physical processes. \cite{Keller2008} did so by assuming only convective core overshooting, while \cite{Neilson2011} proposed a combination of pulsation-driven mass loss and moderate convective core overshooting.  It should be noted that \cite{Keller2008} used a  convective core overshooting parameterization that is a factor of two greater than the parameterization used by \cite{Neilson2011}, which is described in the next section. In fact, we  \citep{Neilson2011} misinterpreted the parameterization used by \cite{Keller2008}, however, this does not affect the result that pulsation-driven mass loss can account for $5-10\%$ of the mass discrepancy, nor the amount of convective core overshooting in that paper.  The large range of measured mass discrepancies and uncertainties determined by \cite{Keller2008} are still consistent with this result. As such, it is useful to define the Cepheid mass discrepancy as the difference between predicted masses from stellar pulsation modeling and predicted masses from stellar evolution models, assuming no convective core overshooting or other extra mixing process or enhanced mass loss.  These stellar evolution models can be referred to as {\it{canonical}} stellar evolution models.  In this way, we can predict a mass difference and then use extra physical processes or parameterizations to predict a resolution to the mass discrepancy.

In their work, \cite{Cassisi2011} made two assumptions about the Cepheid binary: 1) that the Cepheid is evolving along the blue loop, and 2) that the Cepheid's metallicity is $Z=0.008$, with $dy/dZ = 1.4$.  The purpose of this work is to test how the predicted stellar evolution masses depend on these two assumptions and what this suggests about the Cepheid mass discrepancy.  As such, we can verify the results of \cite{Cassisi2011} and test the uniqueness of those results.


\section{Stellar evolution models}
We compute model stellar evolution tracks using the \cite{Yoon2005} stellar evolution code.  The code includes prescriptions for mass loss (for hot stars the \cite{Kudritzki1989} prescription is used while the \cite{dejager1988} prescription is used for cool stars) and convective core overshooting.  Overshooting is treated by assuming that for an evolutionary timescale convective eddies penetrate some fraction of a pressure scale height above the convective core, defined as $d_c = \alpha_c H_P$.  The coefficient $\alpha_c$ is a free input parameter for the evolution code, and is linearly correlated with the amount of Cepheid mass discrepancy \citep{Keller2008}. If one considers a Cepheid with a given luminosity then it also has a specific helium core mass.  The helium core mass is determined by a combination of the initial mass of the star and the parameterized amount of convective core overshooting.  Therefore, if one fits the given luminosity with a canonical stellar model, one would predict a canonical stellar mass.  If the amount of convective overshoot is increased, then a smaller initial mass is required to fit that given luminosity.  Furthermore, one can fit the amount of convective core overshooting to fit a predicted stellar pulsation mass, meaning that the amount of convective core overshooting is correlated to the amount of Cepheid mass discrepancy.

 \cite{Keller2008} argued that to account for a $20\%$ mass discrepancy, stellar evolution models required an average amount of  convective core overshooting,  $\alpha_c =  0.4$ ($\Lambda_c =0.8$ for his models and ranging from $\Lambda_c = 0.1$ to $\Lambda_c = 1.5$).  \cite{Neilson2011} found that a lower amount of convective core overshooting is required, $\alpha_c \approx 0.3$, when pulsation-enhanced mass loss during the Cepheid stage of evolution is included in stellar evolution models.  This result still holds even though \cite{Neilson2011} misinterpreted the overshooting parameterization from \cite{Keller2008}, because of the large range for the measured mass discrepancy. For the case of OGLE-LMC-CEP0227, \cite{Cassisi2011} fit the observed parameters with stellar evolution models assuming an overshooting parameter $\alpha_c = 0.2$, which is consistent with a mass discrepancy of $10\%$, based on the given definition in the Introduction.

We compute stellar evolution models with varying values of metallicity and $\alpha_c$ to fit  the physical parameters describing the two components of the Cepheid binary.  It must be noted that the metallicity and helium abundance in stellar models from the \cite{Heger2000} code are correlated.  From these fits, we determine how robust the \cite{Cassisi2011} results are and if the binary Cepheid might have a mass discrepancy consistent with the LMC Cepheids studied by \cite{Keller2006}.

\section{Cepheid on the first crossing}
Stars evolve across the Cepheid instability strip multiple times; the first crossing occurs when a star evolves along the Hertzsprung Gap to become a red giant star.  This first crossing is usually not considered for most studies of Cepheids \citep[i.e.][]{Bono2000} because the first-crossing timescale is about one-tenth the blue-loop evolution timescale.  Thus, it is a reasonable assumption that most Cepheids are evolving along the blue loop; however, there is only one known eclipsing binary Cepheid pulsating in the fundamental mode (though \cite{Lepischak2004} reported the discovery of a eclipsing binary Cepheid pulsating in the first overtone).  Furthermore, the two stellar components of OGLE-LMC-CEP0227 appear to be at a similar stage of stellar evolution  according to the masses and effective temperatures. Also, because the mass ratio is $q=1.00 \pm 0.01$, one cannot rule out the possibility that both stars might be evolving along the Hertzsprung gap. 

\begin{figure}[t]
\begin{center}
\includegraphics[width=0.47\textwidth]{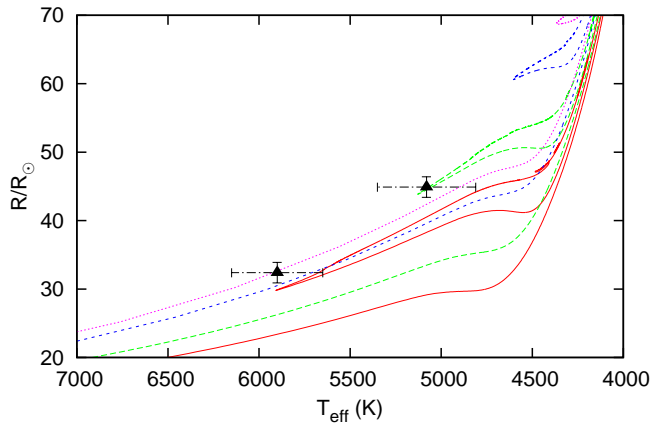}
\end{center}
\caption{Stellar evolution tracks computed using the \cite{Heger2000} stellar evolution program.  The red solid line represents $\alpha_c = 0.2$, green dashed line $\alpha_c = 0.3$, blue small-dashed line $\alpha_c = 0.4$ and the magenta dotted line is $ \alpha_c = 0.45$.  The black triangles with error bars  represent the observed radius and effective temperature of the binary components.}
\label{fig1}
\end{figure}
We compute stellar evolution tracks for stars with initial mass $M=4.21~M_\odot$ with varying amounts of convective core overshooting assumed, $\alpha_c = 0.2, 0.3, 0.4$, and $0.45$.  An initial mass of $4.21~M_\odot$ is chosen so that during post main sequence evolution the model will have a mass of $4.17$ - $4.20~M_\odot$ due to mass loss.  Furthermore, this is the largest value of the stellar mass, consistent with observations, allowing us to predict the minimum amount of convective core overshooting required to agree with the measured stellar parameters. In Fig~\ref{fig1}, we plot evolutionary tracks on the Hertzsprung-Russell diagram, along with the radius and effective temperature for each star in the binary system, where the hotter star is the Cepheid. The stellar models evolve along the Hertzsprung gap with mass $\approx 4.19~M_\odot$  due to mass loss during main sequence evolution.  The stellar evolution tracks suggest that the radii and effective temperatures of the stars in the binary system are consistent with stars evolving across the Hertzsprung gap and a first-crossing Cepheid for $\alpha_c = 0.4$-$0.5$. 

While it is possible for the two stars to be evolving along the Hertzsprung gap, it is highly unlikely.  For instance, we computed stellar evolution models for initial masses $M_1 = 4.22~M_\odot$ and $M_2 = 4.21~M_\odot$ with $\alpha_c = 0.4$  Both evolutionary tracks predict effective temperatures and radii consistent with observed values, but the timescales do not agree.  The more massive evolution model evolves to a temperature of $T_{\rm{eff}} = 5080\pm 270~K$ from the zero age main sequence in about 162~Myr, while the less massive model has an effective temperature $T_{\rm{eff}} = 5900\pm250~K$ in 161~Myr.  However, this timescale difference depend on the difference between the two stellar masses, hence the mass ratio $q$.  Two models with initial masses $4.21$ and $4.22~M_\odot$ has a value of $q = 0.998$, and reducing this mass difference will increase the mass ratio $q \rightarrow 1$ and reduce the timescale differences, all consistent with the measured properties of the binary system.  We can show this in another way,  the primary star in the binary system is more evolved than the secondary, meaning that the primary spent less time evolving on the main sequence, i.e. $\tau \propto M^{2.5}$ \citep{Kippenhahn1994}.  Furthermore, the difference in timescale for the secondary to evolve to the same location as the primary is $\Delta \tau/\tau = 2.5 \Delta M/M$.  The difference in time for the Cepheid to evolve along the Hertzsprung gap to the same location as the primary is of the order $10^5$~yr and the main sequence lifetime is $\tau = 1.60$~Myr, hence $\Delta M/M = 0.0025$ or $q = 0.9975$.  Therefore, for the two stars to be evolving along the Hertzsprung gap with fundamental parameters consistent with the observations for the same age then the mass ratio $q > 0.997$. Only more precise mass determinations of the two stars such that $q$ is sufficiently less than unity will rule out this scenario. 

\section{Cepheid evolution with varying metallicity}
Both \cite{Piet2010} and \cite{Cassisi2011} assume that the system OGLE-LMC-CEP0227 has a metallicity of $Z = 0.008$.  \cite{Keller2006} found, however, LMC Cepheids have metallicities ranging from $Z = 0.005$ to $Z=0.015$. \cite{Romaniello2005, Romaniello2008} also measured a similar, significant metallicity spread in LMC Cepheids.  These works suggest that it is possible that the binary Cepheid may not have the standard Large Magellanic Cloud metallicity. \cite{Cassisi2011} did test stellar evolution models with varying metallicity from $Z=0.006$ to $Z = 0.01$ with $dY/dZ - 1.4$ and found that the change of metallicity did not affect the amount of convective core overshooting necessary to fit the observed radii.

A secondary consideration from Fig.~\ref{fig1} is that the non-Cepheid component has a radius and effective temperature that is consistent with a stellar evolution model with initial mass $4.21~M_\odot$ and $\alpha_c = 0.3$.  The evolution track, however, does not predict a large enough blue loop to compare to the radius and effective temperature of the Cepheid.  It is unclear what determines the width of a blue loop and is very sensitive to the initial parameters of the stellar model, such as convective core overshooting, mass-loss prescription \citep{Neilson2008,Neilson2011}, and helium abundance \citep{Fiorentino2002}.   Even the helium nuclear generation rates affect the blue loop evolution of a star \citep{Morel2010}. In this respect, if we extrapolate the width of the blue loop for the $\alpha_c = 0.3$ evolutionary track to smaller radiii, and hotter effective temperature, then we find that the Cepheid is potentially consistent with $\alpha_c = 0.3$ as well as $\alpha_c  = 0.2$.  Therefore, the binary components may be consistent with blue loop evolution with $\alpha_c >0.2$ for $Z=0.008$.  This result is limited by the large uncertainties of the measured effective temperatures;  more precise measurements of the Cepheid effective temperature would constrain the minimum extent of the Cepheid blue loop and better constrain the input physics for stellar evolution models.

\begin{figure}
\begin{center}
\includegraphics[width=0.47\textwidth]{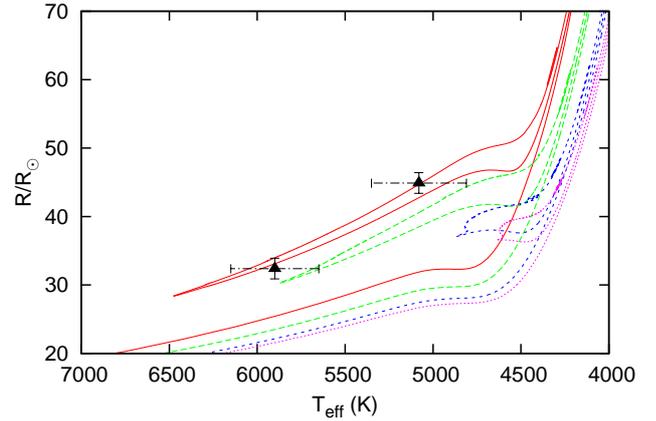}
\end{center}
\caption{Stellar evolution tracks computed using the \cite{Heger2000} stellar evolution program with mass $M=4.21~M_\odot$, convective core overshooting with $\alpha_c = 0.2$ and varying values of the helium abundance, $Y$, and metallicity, $Z$.  The \cite{Heger2000} code assumes the values of $Y$ and $Z$ to be correlated, such that $dY/dZ = 2$.  The red solid line represents $Y,Z=0.250, 0.005$, green dashed line $0.256, 0.008$, blue small-dashed line $0.264, 0.012$ and the magenta dotted line is $0.270, 0.015$.  The black triangles with error bars  represent the observed radius and effective temperature of the binary components.}
\label{fig2}
\end{figure}

In Fig.~\ref{fig2}, we plot stellar evolution tracks with $\alpha_c = 0.2$, $M= 4.21~M_\odot$ with varying metallicity, $Z$, and helium abundance, $Y$.  It is clear that stellar evolution models predict less luminous blue loops for decreasing helium and metal abundances. \cite{Valle2009} found that the blue loop luminosity increased with increased helium abundance for a constant metallicity.  They also found that varying the metallicity for a constant helium abundance changed the extent of the blue loop as well as led to a brighter blue loop for decreasing metallicity. Our models are consistent with this predicted blue loop behavior.

The $M = 4.21~M_\odot$, $\alpha_c = 0.2$ model has a temperature and radius consistent with the measured values for the Cepheid at an age of $\tau = 150~$Myr, while the same model is consistent with the red giant star at an age of $154~$Myr.  Therefore, $\Delta \tau/\tau = 0.026$ hence $q \approx 0.9889$, marginally consistent with the observed mass ratio $q = 1.00\pm0.01$.  Cases where the Cepheid mass is assumed to be smaller will have a greater age for the given Cepheid parameters, likewise models with more convective core overshooting will also have longer main sequence lifetimes and hence larger ages.  Because the ages are greater then the predicted mass ratios for models that agree with the observed parameters will approach unity and agree with the observed $q$.

\begin{figure}
\begin{center}
\includegraphics[width=0.47\textwidth]{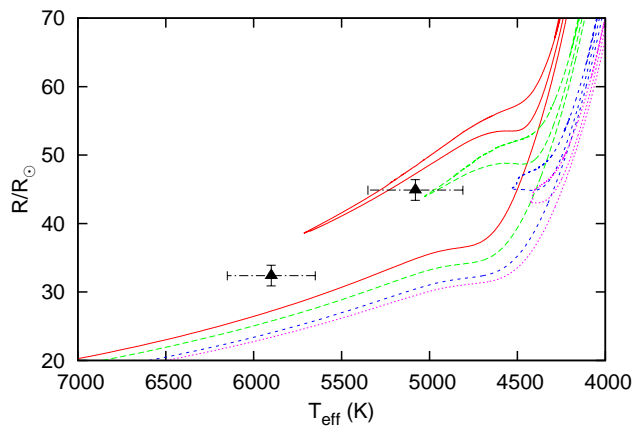}
\end{center}
\caption{Stellar evolution tracks computed using the \cite{Heger2000} stellar evolution program with mass $M=4.14~M_\odot$, convective core overshooting with $\alpha_c = 0.3$ and varying values of the helium abundance, $Y$, and metallicity, $Z$. The line types and colors have the same values of $Y$ and $Z$ as those in Fig.~\ref{fig2}.}
\label{fig3}
\end{figure}

We repeat the analysis with stellar evolution models with mass $M=4.14~M_\odot$ and $\alpha_c = 0.3$ in Fig.~\ref{fig3}. Again, we find that the stellar evolution tracks have blue loops that are too small for fit the observed parameters of the Cepheid.  It is expected that a smaller mass stellar evolution model will have a smaller blue loop luminosity than a more massive model for the same amount of convective core overshooting.  On the other hand, increasing the amount of overshooting in the lower mass model reduces this luminosity difference between the two models.  Therefore, it appears that stellar evolution models with $\alpha_c = 0.3$ and mass $M=4.14~M_\odot$ can fit the observations, assuming standard LMC metallicity and extrapolating the width of the blue loop like before.  The observed mass uncertainties of the  two components of the binary system, while small, still allows for a variation of convective core overshooting.

Furthermore, if the stars are relatively metal-rich, then the observed parameters can be fit with models assuming even larger amounts of convective core overshooting, i.e. smaller masses and smaller metallicities require greater amounts of overshooting to reproduce observed luminosities. Clearly, a metallicity variation consistent with the observed spread of LMC Cepheids suggests that the amount of core overshooting necessary for stellar evolution models to fit the masses of the binary system also varies. Also, noting that a value of $\alpha_c = 0.1$ means a mass discrepancy of $5\%$ then this suggests that the binary system is consistent with a Cepheid mass discrepancy of $10$-$20\%$.

\section{Discussion}
We have presented two alternative scenarios to fit the observed parameters of the Cepheid binary OGLE-LMC-CEP0227. The first scenario assumes the two stars are evolving across Hertzsprung gap, meaning that the Cepheid is on the first crossing of the instability strip. The second scenario considers how fitting the observed radius and effective temperature depends on the assumed metallicity.

If the stars are evolving across the Hertzsprung gap then the observed masses are fit only if $\alpha_c = 0.4$-$0.5$, hence the mass discrepancy would be $20$-$26\%$;  assuming convective core overshooting is the only solution to the mass discrepancy.  If this is the correct scenario describing the evolution of the two stars then one constrains the underlying physics explaining the Cepheid mass discrepancy.  For instance, observations of low-mass binary stars suggest that the convective core overshooting parameter is $\alpha_c = 0.2$ \citep{Sandberg2010, Clausen2010}, leaving about $10-60\%$ mass discrepancy unexplained. Because the Cepheid is on the first crossing pulsation-driven mass loss \citep{Neilson2011} cannot explain the remaining mass discrepancy. The star has not spent enough time as a Cepheid to lose enough mass. Thus rotational mixing remains a viable option \citep{Bono2006, Meynet2009}. However, we note that this scenarios requires a mass ratio $q > 0.997$.

The second scenario assumes that the stars are evolving along the blue loop with non-standard LMC metallicity and helium abundance.  We find that varying the composition plays an important role in predicting the mass of the stars from evolution models, and a change in $Y$ and $Z$, consistent with the spread of metallicities determined for LMC Cepheids, suggests that the amount of convective core overshooting required to match observations is $\alpha_{c}=0.2$-$0.35$. Noting that \cite{Keller2008} and \cite{Neilson2011} argue that a value of $\alpha_c = 0.1$ $(\Lambda_c = 0.2)$ accounts for a mass discrepancy of $5\%$ and if we extrapolate the width of the blue loops, then we have a measure of the mass discrepancy. In this case, we find a Cepheid mass discrepancy of $10-18\%$.  

The evolutionary state of these stars may be understood in greater detail using measurements of the rate of period change, or the nitrogen abundance and by the metallicity in general.  \cite{Turner2006} showed that first-crossing Cepheids have periods that are increasing with time and that the rates of period change tend to be significantly larger than for Cepheids on the third crossing of the instability strip.  Measuring the nitrogen abundance will also provide information on whether the stars have evolved through the first dredge-up on the red giant branch \citep{Hunter2008}. For instance, if a $M= 4.21~M_\odot$ LMC Cepheid is evolving on the first crossing then $[$N/Fe$] \approx  0.3$ while evolving on the blue loop will have $[$N/Fe$]\approx 0.8$.  It must be noted that rotational mixing during main sequence evolution also enhances the nitrogen abundance, implying that first-crossing Cepheids may also have nitrogen enhancements.  However, observations of both the rate of period change and nitrogen abundance would constrain the main sequence rotation and mixing of these stars.  Similarly, \cite{Kovtyukh2005} argued that Cepheids evolving along the blue loop will be lithium depleted and suggested this as another method to constrain the evolution status of Cepheids.  The second scenario can be constrained by spectroscopic observations of the binary stars to determine the metallicity. The helium abundance is much more difficult to determine, helium lines disappear in cool star spectra.  \cite{Dupree2011} measured helium abundance variations in globular cluster stars based on the equivalent width of the He I $\lambda10830$ line, which has been observed in Cepheids \citep{Sasselov1994a, Sasselov1994b}.

It should be noted that \cite{Piet2011} reported the analysis of a second LMC eclipsing binary with a Cepheid component, OGLE-LMC-CEP-1812.  This new Cepheid binary seems to provide another stringent test for binary star evolution, as its mass ratio and evolutionary phases are inconsistent with its age.  The secondary star appears to be about $100~$Myr older than the Cepheid primary.  The Cepheid mass here is also consistent with pulsation masses, but no evolutionary models were tested in that work.

The results presented in this work indicates that there is not enough information known about the binary Cepheid OGLE-LMC-CEP0227 to precisely constrain stellar evolution models and hence resolve the Cepheid mass discrepancy.  The predicted mass discrepancy does not rule out the \cite{Cassisi2011} results, but suggests that other scenarios are possible, such as more convective core overshooting and pulsation-driven mass loss \citep{Neilson2011}.  Only more detailed observations will constrain the underlying physics that solve the discrepancy and test the results of \cite{Cassisi2011} and \cite{Neilson2011}.

\acknowledgements
HRN is grateful for financial support from the Alexander von Humboldt Foundation.  We thank Santi Cassisi for his helpful comments that greatly improved this work.

\bibliographystyle{aa}
\bibliography{cepb}

\begin{thebibliography}{37}
\expandafter\ifx\csname natexlab\endcsname\relax\def\natexlab#1{#1}\fi

\bibitem[{{Bono} {et~al.}(2006){Bono}, {Caputo}, \& {Castellani}}]{Bono2006}
{Bono}, G., {Caputo}, F., \& {Castellani}, V. 2006, \memsai, 77, 207

\bibitem[{{Bono} {et~al.}(2000){Bono}, {Castellani}, \& {Marconi}}]{Bono2000}
{Bono}, G., {Castellani}, V., \& {Marconi}, M. 2000, \apj, 529, 293

\bibitem[{{Caputo} {et~al.}(2005){Caputo}, {Bono}, {Fiorentino}, {Marconi}, \&
  {Musella}}]{Caputo2005}
{Caputo}, F., {Bono}, G., {Fiorentino}, G., {Marconi}, M., \& {Musella}, I.
  2005, \apj, 629, 1021

\bibitem[{{Cassisi} \& {Salaris}(2011)}]{Cassisi2011}
{Cassisi}, S. \& {Salaris}, M. 2011, \apjl, 728, L43

\bibitem[{{Clausen} {et~al.}(2010){Clausen}, {Frandsen}, {Bruntt}, {Olsen},
  {Helt}, {Gregersen}, {Juncher}, \& {Krogstrup}}]{Clausen2010}
{Clausen}, J.~V., {Frandsen}, S., {Bruntt}, H., {et~al.} 2010, \aap, 516, A42

\bibitem[{{Cox}(1980)}]{Cox1980}
{Cox}, A.~N. 1980, \araa, 18, 15

\bibitem[{{de Jager} {et~al.}(1988){de Jager}, {Nieuwenhuijzen}, \& {van der
  Hucht}}]{dejager1988}
{de Jager}, C., {Nieuwenhuijzen}, H., \& {van der Hucht}, K.~A. 1988, \aaps,
  72, 259

\bibitem[{{Dupree} {et~al.}(2011){Dupree}, {Strader}, \& {Smith}}]{Dupree2011}
{Dupree}, A.~K., {Strader}, J., \& {Smith}, G.~H. 2011, \apj, 728, 155

\bibitem[{{Evans} {et~al.}(1997){Evans}, {B\"ohm-Vitense}, {Carpenter},
  {Beck-Winchatz}, \& {Robinson}}]{Evans1997}
{Evans}, N.~R., {B\"ohm-Vitense}, E., {Carpenter}, K., {Beck-Winchatz}, B., \&
  {Robinson}, R. 1997, \pasp, 109, 789

\bibitem[{{Evans} \& {Bolton}(1990)}]{Evans1990}
{Evans}, N.~R. \& {Bolton}, C.~T. 1990, \apj, 356, 630

\bibitem[{{Evans} {et~al.}(2006){Evans}, {Massa}, {Fullerton}, {Sonneborn}, \&
  {Iping}}]{Evans2006}
{Evans}, N.~R., {Massa}, D., {Fullerton}, A., {Sonneborn}, G., \& {Iping}, R.
  2006, \apj, 647, 1387

\bibitem[{{Evans} {et~al.}(2008){Evans}, {Schaefer}, {Bond}, {Bono},
  {Karovska}, {Nelan}, {Sasselov}, \& {Mason}}]{Evans2008}
{Evans}, N.~R., {Schaefer}, G.~H., {Bond}, H.~E., {et~al.} 2008, \aj, 136, 1137

\bibitem[{{Fiorentino} {et~al.}(2002){Fiorentino}, {Caputo}, {Marconi}, \&
  {Musella}}]{Fiorentino2002}
{Fiorentino}, G., {Caputo}, F., {Marconi}, M., \& {Musella}, I. 2002, \apj,
  576, 402

\bibitem[{{Heger} {et~al.}(2000){Heger}, {Langer}, \& {Woosley}}]{Heger2000}
{Heger}, A., {Langer}, N., \& {Woosley}, S.~E. 2000, \apj, 528, 368

\bibitem[{{Hunter} {et~al.}(2008){Hunter}, {Brott}, {Lennon}, {Langer},
  {Dufton}, {Trundle}, {Smartt}, {de Koter}, {Evans}, \& {Ryans}}]{Hunter2008}
{Hunter}, I., {Brott}, I., {Lennon}, D.~J., {et~al.} 2008, \apjl, 676, L29

\bibitem[{{Keller}(2008)}]{Keller2008}
{Keller}, S.~C. 2008, \apj, 677, 483

\bibitem[{{Keller} \& {Wood}(2006)}]{Keller2006}
{Keller}, S.~C. \& {Wood}, P.~R. 2006, \apj, 642, 834

\bibitem[{{Kippenhahn} \& {Weigert}(1994)}]{Kippenhahn1994}
{Kippenhahn}, R. \& {Weigert}, A. 1994, {Stellar Structure and Evolution}
  ({Berlin}: {Springer-Verlag})

\bibitem[{{Kovtyukh} {et~al.}(2005){Kovtyukh}, {Wallerstein}, \&
  {Andrievsky}}]{Kovtyukh2005}
{Kovtyukh}, V.~V., {Wallerstein}, G., \& {Andrievsky}, S.~M. 2005, \pasp, 117,
  1182

\bibitem[{{Kudritzki} {et~al.}(1989){Kudritzki}, {Pauldrach}, {Puls}, \&
  {Abbott}}]{Kudritzki1989}
{Kudritzki}, R.~P., {Pauldrach}, A., {Puls}, J., \& {Abbott}, D.~C. 1989, \aap,
  219, 205

\bibitem[{{Lepischak} \& {Welch}(2004)}]{Lepischak2004}
{Lepischak}, D. \& {Welch}, D.~L. 2004, in IAU Colloq. 193, Vol. 310, Variable
  Stars in the Local Group, ed. {D.~W.~Kurtz \& K.~R.~Pollard}, ASP Conf. Ser.,
  372

\bibitem[{{Lovekin} \& {Goupil}(2010)}]{Lovekin2010}
{Lovekin}, C.~C. \& {Goupil}, M.-J. 2010, \aap, 515, A58

\bibitem[{{Meynet}(2009)}]{Meynet2009}
{Meynet}, G. 2009, in Lecture Notes in Physics, Vol. 765, The Rotation of Sun
  and Stars ({Berlin}: {Springer Verlag}), 139--169

\bibitem[{{Morel} {et~al.}(2010){Morel}, {Provost}, {Pichon}, {Lebreton}, \&
  {Th{\'e}venin}}]{Morel2010}
{Morel}, P., {Provost}, J., {Pichon}, B., {Lebreton}, Y., \& {Th{\'e}venin}, F.
  2010, \aap, 520, A41

\bibitem[{{Neilson} {et~al.}(2011){Neilson}, {Cantiello}, \&
  {Langer}}]{Neilson2011}
{Neilson}, H.~R., {Cantiello}, M., \& {Langer}, N. 2011, \aap, 529, L9

\bibitem[{{Neilson} \& {Lester}(2008)}]{Neilson2008}
{Neilson}, H.~R. \& {Lester}, J.~B. 2008, \apj, 684, 569

\bibitem[{{Pietrzynski} {et~al.}(2011){Pietrzynski}, {Thompson}, {Graczyk.},
  {Gieren}, {Pilecki}, {Udalski}, {Soszynski}, {Bono}, {Konorski}, {Nardetto},
  \& {Storm}}]{Piet2011}
{Pietrzynski}, G., {Thompson}, I., {Graczyk.}, D., {et~al.} 2011,
  ArXiv:1109.5414

\bibitem[{{Pietrzy{\'n}ski} {et~al.}(2010){Pietrzy{\'n}ski}, {Thompson},
  {Gieren}, {Graczyk}, {Bono}, {Udalski}, {Soszy{\'n}ski}, {Minniti}, \&
  {Pilecki}}]{Piet2010}
{Pietrzy{\'n}ski}, G., {Thompson}, I.~B., {Gieren}, W., {et~al.} 2010, \nat,
  468, 542

\bibitem[{{Romaniello} {et~al.}(2005){Romaniello}, {Primas}, {Mottini},
  {Groenewegen}, {Bono}, \& {Fran{\c c}ois}}]{Romaniello2005}
{Romaniello}, M., {Primas}, F., {Mottini}, M., {et~al.} 2005, \aap, 429, L37

\bibitem[{{Romaniello} {et~al.}(2008){Romaniello}, {Primas}, {Mottini},
  {Pedicelli}, {Lemasle}, {Bono}, {Fran{\c c}ois}, {Groenewegen}, \&
  {Laney}}]{Romaniello2008}
{Romaniello}, M., {Primas}, F., {Mottini}, M., {et~al.} 2008, \aap, 488, 731

\bibitem[{{Sandberg Lacy} {et~al.}(2010){Sandberg Lacy}, {Torres}, {Claret},
  {Charbonneau}, {O'Donovan}, \& {Mandushev}}]{Sandberg2010}
{Sandberg Lacy}, C.~H., {Torres}, G., {Claret}, A., {et~al.} 2010, \aj, 139,
  2347

\bibitem[{{Sasselov} \& {Lester}(1994{\natexlab{a}})}]{Sasselov1994a}
{Sasselov}, D.~D. \& {Lester}, J.~B. 1994{\natexlab{a}}, \apj, 423, 777

\bibitem[{{Sasselov} \& {Lester}(1994{\natexlab{b}})}]{Sasselov1994b}
{Sasselov}, D.~D. \& {Lester}, J.~B. 1994{\natexlab{b}}, \apj, 423, 785

\bibitem[{{Stobie}(1969)}]{Stobie1969}
{Stobie}, R.~S. 1969, \mnras, 144, 511

\bibitem[{{Turner} {et~al.}(2006){Turner}, {Abdel-Sabour Abdel-Latif}, \&
  {Berdnikov}}]{Turner2006}
{Turner}, D.~G., {Abdel-Sabour Abdel-Latif}, M., \& {Berdnikov}, L.~N. 2006,
  \pasp, 118, 410

\bibitem[{{Valle} {et~al.}(2009){Valle}, {Marconi}, {Degl'Innocenti}, \& {Prada
  Moroni}}]{Valle2009}
{Valle}, G., {Marconi}, M., {Degl'Innocenti}, S., \& {Prada Moroni}, P.~G.
  2009, \aap, 507, 1541

\bibitem[{{Yoon} \& {Langer}(2005)}]{Yoon2005}
{Yoon}, S.-C. \& {Langer}, N. 2005, \aap, 443, 643

\end{thebibliography}

\end{document}